# Coupling SPH and thermochemical models of planets: Methodology and example of a Mars-sized body

G.J. Golabek[1,2,*], A. Emsenhuber[3], M. Jutzi[3], E.I. Asphaug[4,5] & T.V. Gerya[2]

[1] Bayerisches Geoinstitut, University of Bayreuth, Universitätsstrasse 30, 95440 Bayreuth, Germany.
[2] ETH Zurich, Institute of Geophysics, Sonneggstrasse 5, 8092 Zürich, Switzerland.
[3] University of Bern, Physics Institute, Space Research and Planetary Sciences, Center for Space and Habitability, Gesellschaftsstrasse 6, 3012 Bern, Switzerland.
[4] Arizona State University, School of Earth and Space Exploration, PO Box 876004, Tempe, AZ 85287, USA.
[5] University of Arizona, Lunar and Planetary Laboratory, 1269 E. University Blvd, Tucson, AZ 85721, USA.

* Corresponding author. G.J. Golabek (gregor.golabek@uni-bayreuth.de)

**Abstract**

Giant impacts have been suggested to explain various characteristics of terrestrial planets and their moons. However, so far in most models only the immediate effects of the collisions have been considered, while the long-term interior evolution of the impacted planets was not studied. Here we present a new approach, combining 3-D shock physics collision calculations with 3-D thermochemical interior evolution models.

We apply the combined methods to a demonstration example of a giant impact on a Mars-sized body, using typical collisional parameters from previous studies. While the material parameters (equation of state, rheology model) used in the impact simulations can have some effect on the long-term evolution, we find that the impact angle is the most crucial parameter for the resulting spatial distribution of the newly formed crust. The results indicate that a dichotomous crustal pattern can form after a head-on collision, while this is not the case when considering a more likely grazing collision. Our results underline that end-to-end 3-D calculations of the entire process are required to study in the future the effects of large-scale impacts on the evolution of planetary interiors.

# 1. Introduction

Towards the end of terrestrial planetary accretion, giant impacts are inevitable [*Wetherill, 1985; Melosh, 1990; Canup & Asphaug, 2001*]. It has been suggested that they can explain various characteristics of the terrestrial planets like Mercury's anomalously thin silicate mantle [*Benz et al., 1988, 2007; Asphaug and Reufer, 2014*], the retrograde spin of Venus [*Alemi and Stevenson, 2006*], the origin of the Earth's Moon [*Hartmann and Davis, 1975; Cameron and Ward, 1976; Canup and Asphaug, 2001; Canup, 2004, 2012; Ćuk and Stewart, 2012; Reufer et al., 2012*] and the formation of the martian dichotomy [*Wilhelms and Squyres, 1984; Frey and Schultz, 1988; Andrews-Hanna et al., 2008; Nimmo et al., 2008; Marinova et al., 2008, 2011*]. For the case of Mars, numerical model results suggest that the northern lowlands of Mars could correspond to a giant impact basin formed after primordial crust formation. However in these models only the impact process and its immediate aftermath are considered, while the longer-term evolution featuring post-collision crust formation is not studied.

Thermochemical codes consider the physical evolution of the planetary interior, and are better suited than impact models to study the long-term evolution of planetary interiors following a major collision. For the case of the martian dichotomy these endogenic models [e.g., *Weinstein, 1995*], suggest a degree-1 mantle upwelling underneath the southern highlands [*Zhong and Zuber, 2001; Roberts and Zhong, 2006; Zhong, 2009; Keller and Tackley, 2009; Šrámek and Zhong, 2010, 2012*]. These models rely on a high Rayleigh number and a particular viscosity profile to form a low degree convective planform within the geological constraints for Mars dichotomy formation. Also for the case of the martian dichotomy formation, a hybrid exogenic-endogenic approach was suggested [*Reese and Solomatov, 2006, 2010; Reese et al., 2010; Golabek et al., 2011*], proposing an impact-induced regional magma ocean and subsequent superplume in the southern hemisphere.

These thermochemical simulations usually rely on highly simplified impact models, for instance a uniform or localized deposition of energy or matter representative of the collision [e.g. *Wilhelms and Squyres, 1984*]. Although a simplified approach might be suitable for the limited case of a head-on impact, it is well established that impacts at 45 degrees are most likely [*Gilbert, 1893; Shoemaker, 1962*]. This is true for cratering and for giant impacts, however for cratering, the infinite target means that impact angle does not matter very much. No matter what the angle, the projectile can not find its way beyond the target, and there is no abrupt transition to 'grazing' behavior until impact angles of around 75 degrees from vertical [*Melosh, 1989*]. However for giant impacts and other similar-sized collisions [*Asphaug, 2010*], even a close-to-head-on impact (less than 20 degrees) can allow for lopsided evolution of material. Thus, the combined effects of gravity, angular momentum, and linear downrange momentum require 3-D models.

Thus to better understand the influence of giant impacts on the early evolution of terrestrial planets, methodological improvements are necessary, ideally using the advantages of both methods to overcome the current model limitations. In a recent study, results of 2-D axisymmetric impact calculations were used as input in thermochemical evolution models [*Rolf et al., 2017*]. We present here a new approach using the results of 3-D large-scale collision models performed with a shock physics code to be used as initial conditions in a thermochemical code and present demonstration calculations.

In section 2 we introduce the methods used by both codes, one for giant impacts and one for the thermochemical evolution. In section 3 we describe the model scenario of our demonstration calculation, which we use to test the coupled method. In section 4 we discuss the effect of parameters like the equation of state, the rheology and the transfer time on the post-impact crust formation. The two final sections are devoted to discussion and outlook.

**2. Methodology**

*2.1 SPH code*

We use a smoothed particle hydrodynamics (SPH) code [*Benz and Asphaug, 1994; Reufer et al. 2012; Jutzi et al., 2013*] to model the collision. This code is based on the SPHLATCH code developed by *Reufer et al. [2012]* and includes self-gravity, a newly implemented strength model and various equations of state (EoS). The details of the strength and EoS models are described in *Emsenhuber et al., in press*. Here we only give a short summary of the basic method.

SPH uses a Lagrangian representation where material is divided into particles. Quantities are interpolated ('smoothed') over a certain length by summing over surrounding particles (called neighbours) using a kernel function

$$B(\vec{x}) = \sum_i B_i W(\vec{x} - \vec{x}_i, h_i) \quad (1)$$

where $\vec{x}_i$, $h_i$ and $V_i$ are the position, smoothing length, and volume of particle $i$, respectively. $B_i$ represents the quantity (field variable) to be interpolated and $W(\vec{x}, h)$ is the kernel. This interpolation scheme is used in SPH to solve the relevant differential equations [see *Emsenhuber et al., in press*]. SPH is an adaptive resolution technique, thus large volumes of space can be modelled economically. Indeed 'space' has no particles and costs no resolution. The method is therefore quite popular in three-dimensional (3-D) planetary collisional studies [e.g. *Jutzi et al., 2015*], and includes good equations of state and treatments of shock and self-gravity. But being a 2$^{nd}$ order method limited to the Courant timescale (that is, the sound-crossing time of a particle smoothing length), it is not suited to model the longer-term post-impact evolution.

Body accelerations are the result of the pressure gradient, and for this SPH uses the pressure as computed from an equation of state (EoS; see below) which is a function $P(\rho,u)$ of density $\rho$ and internal energy $u$. For solid materials the pressure gradient is generalized into a stress tensor [*Benz and Asphaug, 1994*]. We use a Drucker-Prager-like yield criterion [*Collins et al., 2004, Jutzi, 2015*] with yield strength $\sigma_i$ of intact material, which is a function of temperature [*Collins et al., 2004*]. A more sophisticated model for geological materials, which includes also a tensile fracture and a porosity model, and its application in SPH (using a different code) is described in *Jutzi [2015]*; for problems at the global-scale of a Mars-sized body these aspects can be ignored.

As for the EoS we use either Tillotson [*Tillotson, 1962*] or ANEOS [*Thompson and Lauson, 1972; Melosh, 2007*]. The Tillotson equation of state provides both the pressure $P$ and the speed of sound $v_s$ as output, but it does not give the temperature $T$, so as an approximation we use the internal energy $u$ as a proxy for temperature by dividing by the heat capacity $c_p$. However it should be noted that eq. (2) does not take into account that heat capacity is temperature-dependent.

$$T = u/c_p \quad (2)$$

For a more physical computation of temperature, and to allow for phase transitions, which do not occur in the Tillotson EoS, we use ANEOS [*Thompson and Lauson, 1972*] for iron and M-ANEOS [*Melosh, 2007*] for silicates. These equations of state provide numerous output variables including the temperature and phase information (e.g. melt and vapor fraction). Depending on the parameters used in the equation of state, this phase information can be given in different ways. For iron, it can be used to infer the melting temperature at a given pressure. However, this is not the case for silicates, for which we apply the same procedure as used in [*Senft and Stewart, 2009*] to obtain the melting temperature at a given pressure.

Numerous tests have been made comparing SPH with other codes in the area of impact contact and compression and the early stages of dynamical evolution. *Canup et al. [2013]* find that SPH and CTH (a popular grid-based shock physics hydrocode) both give similar answers for the amount of mass and momentum ejected into a protolunar disk for a Moon-forming giant impact scenario, although varying in detailed aspects such as clumping. *Pierazzo et al. [2008]* make comparisons and find overall agreement for the early impact ejection phase of large-scale impact cratering calculations.

*2.2 I3ELVIS code*

The study of giant impacts over the past 30 years has depended on hydrocode methods such as SPH, whose methods are generally familiar to the planetary science community. Global thermochemical methods are required to study the post-collision evolution of a planetary object after a giant impact, so we describe that methodology in more detail.

We consider 3-D creeping flow using the Boussinesq approximation, in which both thermal and chemical buoyancy forces are included. For this purpose we performed 3-D simulations using the code I3ELVIS by applying the 'spherical-Cartesian' methodology [see also *Gerya and Yuen, 2007*]. The methodology combines finite differences on a fully staggered rectangular Eulerian grid with a Lagrangian marker-in-cell technique for solving the momentum, continuity and temperature equations. In order to model a planetary body on a Cartesian grid, the code calculates self-consistently the gravitational field for a self-gravitating planetary body.

In detail, the gravitational potential can be described using the Poisson equation

$$\frac{\partial^2 \phi}{\partial x^2} + \frac{\partial^2 \phi}{\partial y^2} + \frac{\partial^2 \phi}{\partial z^2} = 4\pi G \rho \quad (3)$$

where $\phi$ is the gravitational potential and $G$ is the gravitational constant. Knowing the gravitational potential, the components of the gravity vector can be computed as follows:

$$g_x = -\frac{\partial \phi}{\partial x}; \quad g_y = -\frac{\partial \phi}{\partial y}; \quad g_z = -\frac{\partial \phi}{\partial z} \quad (4)$$

We assume the Boussinesq approximation in 3-D geometry:

$$\frac{\partial v_x}{\partial x} + \frac{\partial v_y}{\partial y} + \frac{\partial v_z}{\partial z} = 0 \quad (5)$$

where $v_x$, $v_y$ and $v_z$ are the $x$, $y$ and $z$ component of the velocity vector, respectively. The local density $\rho$ depends explicitly on temperature $T$, pressure $P$, composition $c$ and melt fraction $\varphi$.

The density of iron and both solid and molten silicates varies with $P$-$T$ conditions according to the relation:

$$\rho = \rho_0 [1 - \alpha(T - T_0)] \cdot [1 + \beta(P - P_0)] \quad (6)$$

where $\rho_0$ is the density of the distinct material at $T_0 = 298.15$ K and $P_0 = 10^5$ Pa, $\alpha$ is the thermal expansion coefficient and $\beta$ is the compressibility coefficient.

For partially molten silicates we take additionally into account that the density will vary with the melt fraction $\varphi$ as follows [see *Burg and Gerya, 2005*]:

$$\rho_{eff} = \rho_{Si-sol} - \varphi[\rho_{Si-sol} - \rho_{Si-liq}] \quad (7)$$

where $\rho_{Si-sol}$ and $\rho_{Si-liq}$ are the pressure and temperature-dependent densities of the solid silicates and liquid silicates, respectively.

For this purpose we use a batch-melting model assuming a peridotite composition. Pressure-dependent parameterizations for both the solidus and liquidus temperatures, $T_{sol}$ and $T_{liq}$ of peridotite are applied [*Herzberg et al. 2000; Trønnes and Frost, 2002; Wade and Wood 2005*]:

$T_{sol}$ [K] = 1416.2 + 58.3 $P$ [GPa] + 52.3 $P^2$ [GPa] - 16.3 $P^3$ [GPa] + 2.3 $P^4$ [GPa] - 0.2 $P^5$ [GPa] + 8·10$^{-3}$ $P^6$ [GPa] - 2·10$^{-4}$ $P^7$ [GPa] + 2·10$^{-6}$ $P^8$ [GPa] (8)

$T_{liq}$ [K] = 1973 + 28.57 $P$ [GPa] (9)

For $T \leq T_{sol}$, the silicate melt fraction $\varphi$ is zero, for $T \geq T_{liq}$, it is equal to 1. In the intermediate temperature range $T_{sol} < T < T_{liq}$, the melt fraction is assumed to increase linearly with temperature according to the following relation [*Burg and Gerya, 2005*]:

$$\varphi = \frac{T - T_{sol}}{T_{liq} - T_{sol}} \quad (10)$$

Both consumption and release of latent heat due to melting and freezing of silicates are taken into account.

The 3-D Stokes equations for creeping flow take the form:

$$\frac{\partial \sigma_{xx}}{\partial x} + \frac{\partial \sigma_{xy}}{\partial y} + \frac{\partial \sigma_{xz}}{\partial z} - \frac{\partial P}{\partial x} = -g_x \rho \quad (11)$$

$$\frac{\partial \sigma_{yx}}{\partial x} + \frac{\partial \sigma_{yy}}{\partial y} + \frac{\partial \sigma_{yz}}{\partial z} - \frac{\partial P}{\partial y} = -g_y \rho \quad (12)$$

$$\frac{\partial \sigma_{zx}}{\partial x} + \frac{\partial \sigma_{zy}}{\partial y} + \frac{\partial \sigma_{zz}}{\partial z} - \frac{\partial P}{\partial z} = -g_z \rho \quad (13)$$

where $\sigma_{ij}$ represents the components of the deviatoric stress tensor and $P$ is the total pressure, which includes both dynamic and lithostatic components.

We have adopted a Lagrangian frame in which the energy conservation equation takes the following form [*Gerya and Yuen, 2003, 2007*]:

$$\rho c_p \left(\frac{DT}{Dt}\right) = -\frac{\partial q_x}{\partial x} - \frac{\partial q_y}{\partial y} - \frac{\partial q_z}{\partial z} + H_r + H_s + H_L \quad (14)$$

where D/D$t$ is the substantive time derivative, $H_r$ is the radiogenic heating term, $H_s$ is the shear heating term, $H_L$ is the latent heating term and $q_i$ is a heat flux component.

The shear heating term is given by

$$H_s = \sigma_{xx}\dot{\varepsilon}_{xx} + \sigma_{yy}\dot{\varepsilon}_{yy} + \sigma_{zz}\dot{\varepsilon}_{zz} + 2\sigma_{xy}\dot{\varepsilon}_{xy} + 2\sigma_{xz}\dot{\varepsilon}_{xz} + 2\sigma_{yz}\dot{\varepsilon}_{yz} \quad (15)$$

where $\dot{\varepsilon}_{ij}$ are the components of the strain-rate tensor defined as

$$\dot{\varepsilon}_{ij} = \frac{1}{2}\left(\frac{\partial v_i}{\partial x_j} + \frac{\partial v_j}{\partial x_i}\right) \quad (16)$$

The heat flux $q_i$ is defined as

$$q_x = -k\frac{\partial T}{\partial x}, q_y = -k\frac{\partial T}{\partial y}, q_z = -k\frac{\partial T}{\partial z} \quad (17)$$

where $k$ is the thermal conductivity.

We employ a viscous constitutive relationship between stress and strain-rate with $\eta$ representing the effective viscosity:

$$\sigma_{ij} = 2\eta\dot{\varepsilon}_{ij} \quad (18)$$

We use for solid silicates a viscosity $\eta$, which depends on temperature $T$, pressure $P$ and strain rate $\dot{\varepsilon}$ defined in terms of deformation invariants [*Ranalli, 1995*] as:

$$\eta_{eff} = A^{1/n} exp\left(\frac{E_a + PV_a}{nRT}\right) \dot{\varepsilon}_{II}^{(1-n)/n} \quad (19)$$

where $\dot{\varepsilon}_{II} = \sqrt{\frac{1}{2}\dot{\varepsilon}_{ij}\dot{\varepsilon}_{ij}}$ is the second invariant of the strain-rate tensor and $A$, $E_a$ and $n$ are the pre-exponential parameter, the activation energy and the power law coefficient, respectively. $R$ is the gas constant and $V_a$ is the activation volume.

For solid silicates, this ductile rheology is combined with a brittle rheology to yield an effective viscoplastic rheology. For this purpose the Mohr-Coulomb [e.g., *Ranalli, 1995*] and Peierls [e.g., *Kameyama et al., 1999*] yield criteria are simultaneously implemented by limiting the creep viscosity as follows:

$$\eta_{eff} \leq \frac{C + Pf}{2\dot{\varepsilon}_{II}} \quad (20)$$

and for $C + Pf \geq \sigma_P$

$$\eta_{eff} \leq \frac{\sigma_P}{2\dot{\varepsilon}_{II}} \quad (21)$$

where $C$ is the cohesion, $f$ is the internal friction coefficient and $\sigma_P$ is the Peierls stress limit.

The presence of a melt fraction $\varphi > 0.1$ has for silicate material an additional influence on the effective viscosity as given here [*Pinkerton and Stevenson, 1992*]

$$\eta_{eff} = \eta_{Si-liq} exp\left\{\left[2.5 + \left(\frac{1-\varphi}{\varphi}\right)^{0.48}\right](1-\varphi)\right\} \quad (22)$$

where $\eta_{Si-liq}$ is the viscosity of molten silicates, which we assume to be a constant.

*2.2.1 Effective thermal conductivity of largely molten silicates:*

Within a narrow silicate melt fraction range, the silicate behaviour undergoes a transition from a solid-like material to a low viscosity crystal suspension [see *Costa et al., 2009; Solomatov, 2015*]. Thus, the viscosity of largely molten silicates is $\eta_{Si-liq} \sim 10^{-4}\text{-}10^{2}$ Pa s [*Rubie et al., 2003; Liebske et al., 2005; Bottinga and Weill, 1972*]. Hence, due to the low viscosity, both the Rayleigh *Ra* and the Nusselt number *Nu* will be high and cooling will be a very efficient process.

Due to numerical restrictions, the lower cut-off viscosity in the numerical model is $\eta_{num} = 10^{18}$ Pa s, many orders of magnitude higher than realistic viscosities for largely molten silicates. It was suggested that the heat flux from a magma ocean [e.g. *Solomatov, 2015*] can be described using the so-called soft turbulence model [*Kraichnan, 1962; Siggia, 1994*]. In this model the expected convective heat flux *q* is given as:

$$q = 0.089 \frac{k(T-T_{surf})}{L} Ra^{1/3} \quad (23)$$

where the Rayleigh number *Ra* is defined as

$$Ra = \frac{\alpha g (T-T_{surf}) \rho_{eff}^2 c_P D^3}{k \eta_{Si-liq}} \quad (24)$$

$T_{surf}$ is the surface temperature and *D* is the depth of the magma ocean.

Depending on the actual silicate melt viscosity in the numerical model $\eta_{num}$, one can estimate an increased effective thermal conductivity $k_{eff}$ by using the theoretically expected heat flux from a low viscosity magma ocean *q* from eq. (23). This effective thermal conductivity can simulate the heat flux of a medium with a realistic magma ocean viscosity [*Zahnle et al., 1988; Tackley et al., 2001; Hevey and Sanders, 2006; Golabek et al., 2011, 2014*], despite our numerical limitations. Combining eq. (23) and (24) this can be done using the following expression for the effective thermal conductivity $k_{eff}$:

$$k_{eff} = \left(\frac{q}{0.089}\right)^{\frac{3}{2}} \frac{1}{(T-T_{surf})^2 \rho_{eff}} \left(\frac{\alpha g c_P}{\eta_{num}}\right)^{-\frac{1}{2}} \quad (25)$$

*2.2.2 Crust formation*

Partial melting of the mantle, melt extraction and percolation toward the bottom of the forming basaltic crust is implemented in a simplified manner. According to our model, mafic magma added to the crust is balanced by melt production and extraction in the mantle. However, melt percolation is not modelled directly and is considered to be nearly instantaneous. The standard (i.e. without melt extraction) volumetric degree of mantle melting *φ* changes with pressure and temperature according to the linear batch melting model (see eq. 10). Lagrangian markers track the amount of melt extracted during the evolution of each numerical experiment. The total amount of melt, $\varphi_{tot}$ for every marker takes into account the amount of previously extracted melt and is calculated as

$$\varphi_{tot} = \varphi - \sum_N \varphi_{ext} \quad (26)$$

where $\sum_N \varphi_{ext}$ is the total melt fraction extracted during the previous *N* extraction episodes. The rock is considered to be non-molten (refractory) when the extracted melt fraction is larger than the current one (i.e. when $\sum_N \varphi_{ext} > \varphi$). Since the extracted melt fraction propagates much faster than the rocks deform [*Condomines et al., 1988*], melts produced at depth are moved towards the surface and added to the bottom of the forming crust. In order to ensure melt volume conservation and account for mantle compaction and subsidence in response to the melt extraction, melt addition to the bottom of the crust is performed at every time step by converting the shallowest markers of mantle into crustal markers. The local volume of these new crustal markers matches the local volume of extracted melt computed for the time step. Basaltic melts are assumed to be only extracted from relatively shallow (≤300 km depth) mantle regions with low degree of melting ($\varphi \leq 0.2$). This corresponds roughly to the pyroxene fraction of a fertile mantle, being the main component in the basaltic to andesitic composition of crustal material [*McKenzie and Bickle, 1988*]. For simplicity, we do not take partitioning of heat producing elements into the crustal material into account.

## 3. Model setup

We now apply the combined methods to the example of a giant impact on a Mars-sized body, using typical collisional parameters from previous studies. We perform a series of 8 SPH collision models, using a Mars-mass target body ($R = 3436$ km) and an impactor with 1000 km radius made of silicates and iron, with no initial crust on both bodies. The collision occurs at 5 Myr after CAI formation, within the constraints for the last giant impact onto a smaller planetary embryo [*Nimmo & Kleine, 2007*; *Mezger et al. 2013*; *Morishima et al. 2013*]. For each combination of physical properties (solid and fluid rheologies with either Tillotson or ANEOS EoS) we test different impact angles of 0° (head-on collision) and 45° (grazing collision), as defined in [*Asphaug, 2010*]. For details see Table 1. We use in all our models a collision at mutual escape velocity [*Asphaug, 2010*]. With a value ~$3 \cdot 10^{-3}$ for the ratio of kinetic impactor energy over gravitational binding energy, this model setup is expected to cause significant post-impact melting [*Reese and Solomatov, 2006*], resulting in significant crust formation following the giant impact. The simulations used here are a subset of the suite of calculations performed in [*Emsenhuber et al., in press*].

The target and impactor both start with a Mars-like internal structure with an iron core radius half of the body radius. For simulations using Tillotson EoS, we start with an isothermal mantle with an initial temperature close to the surface solidus temperature of peridotite ($T_{Si} = 1500$ K) [*Reese and Solomatov, 2010*]. The core temperature is initially set for both objects to a constant value of $T_{Fe} = 1800$ K, which ensures that at the pressure conditions at the center of the target body the assumed eutectic Fe-FeS is molten [*Chudinkovich and Boehler, 2007*]. For the simulations with ANEOS, an isentropic profile is used for which the temperature at the center of the core is 1800 K and 1500 K for the silicate at the core-mantle boundary.

We begin setting up self-gravitating, hydrostatically-equilibrated planetary objects, starting with one dimensional (1-D) spherically symmetric bodies modelled using a Lagrangian hydrocode [*Benz, 1991*], with the same EoS model as for SPH. This profile is evolved by computing the force balance between self-gravity and pressure (including a damping term), until hydrostatic equilibrium is reached so that radial velocities are small (less than 1% of the escape velocity). Afterwards we transfer the 1-D radial profile onto SPH particles that are placed onto a 3-D lattice. Parameters of each particle are copied from the 1-D profile according to the radius. As particles are equally spaced on the lattice, variation of density is taken into account by adjusting the particle mass. The spherical SPH bodies are then also evolved in a last initializing step to reach hydrostatic equilibrium and negligible radial velocities. Thus the SPH simulations start with two relaxed, differentiated spherical planetary objects. To model the collision they start at an initial distance of several radii, so that during the approach they begin to deform tidally prior to the collision, which has an effect on the collision outcome.

The SPH simulations are performed with a resolution of about one million particles for the target. The number of particles for the impactor is scaled according to the mass ratio between the two planetary objects, so that particle spacing *h* is approximately constant, for the greatest numerical accuracy during the collision. The corresponding smoothing length is then approximately 60 km for both bodies. For more details of the SPH calculations see *Emsenhuber et al., in press*.

The results from SPH are transferred to I3ELVIS at earliest after the collision when (i) the seismic shock waves in the target body have decayed and (ii) after the bulk of ejecta material has either fallen back to the surface of the target body or has escaped (see section 4.1.2). For most cases the transfer time is 18 hours after the collision, several times the self-gravity timescale. To study the sensitivity of the post-collision model evolution to the transfer time, we performed for the reference case (ANEOS EoS, grazing collision) additional thermochemical calculations starting at 8 and 12 hours after the collision (see also Table 1).

The temperature and composition results are interpolated (using the SPH Kernel function as in eq. (1)) onto a Eulerian equally spaced grid of $(10,000$ km$)^3$ size centered on the middle of the post-impact target body. The resulting grid spacing is 20 km, which corresponds to a grid resolution of $501^3$. The computational domain is filled with $5.24 \cdot 10^8$ randomly distributed Lagrangian markers. The informations are then transferred from these nodal points to the markers using a bilinear interpolation scheme described in detail elsewhere [*Gerya and Yuen, 2003, 2007*]. To avoid unrealistic silicate melt densities, we introduce in the I3ELVIS models a lower density cut-off at 2500 kg/m$^3$.

The I3ELVIS model requires some cleanup of the SPH input. Thus material at a distance of more than 3700 km from the planetary center is considered to be remnant ejecta material and is therefore deleted. The rest of the box is filled with sticky air material of nearly zero density, constant viscosity ($\eta_{sa} = 10^{18}$ Pa s) and constant temperature of $T_{sa} = 220$ K. This layer represents an infinite reservoir to absorb heat released from the planetary body [*Golabek et al., 2014; Tkalcec et al., 2013*] and ensures a free surface of the planetary body [*Schmeling et al., 2008; Crameri et al., 2012*].

For numerical reasons, the viscosity range is limited to six orders of magnitude. The silicate viscosity is cut off at lower and upper limits of $10^{18}$ and $10^{24}$ Pa s, whereas the iron viscosity is kept constant at $10^{18}$ Pa s. A similar value for the maximum silicate viscosity ($10^{24}$ - $10^{28}$ Pa s) was suggested before [*Karato and Murthy, 1997*]. It should be noted that the iron viscosity in the model ($\eta_{Fe} = 10^{18}$ Pa s) is, for numerical reasons, many orders of magnitude higher than its physical value, which is expected to range from $10^{-2}$ Pa s (in the liquid state) [*Rubie et al., 2015*] up to $10^{12}$ Pa s in the solid state under high temperature and pressure conditions [*Yunker and Van Orman, 2007*].

For all models, we consider time-dependent radioactive heating by both short- ($^{26}$Al, $^{60}$Fe) and long-lived ($^{40}$K, $^{235}$U, $^{238}$U, $^{232}$Th) radiogenic isotopes. In the early solar system, $^{26}$Al is by far the dominant radioactive energy source and the initial $^{26}$Al/$^{27}$Al is taken as 5.85· 10$^{-5}$ [*Thrane et al., 2006*], this value representing an upper limit for the abundance of $^{26}$Al [see *Jacobsen et al., 2008*].

We use a wet olivine rheology [*Ranalli, 1995*]. This is reasonable as olivine represents the majority of the martian mantle composition and is weaker relative to pyroxenes, thus controlling mantle deformation [*Mackwell, 1991*]. The Peierls stress limit $\sigma_P$ employed for wet olivine rheology [*Katayama and Karato, 2008*] is 2.9 GPa. As suggested for a Mars-like mantle composition, we apply a density crossover between solid and molten silicates at 600 km depth [*Suzuki et al., 1998*]. For all other physical parameters employed see Table 2.

## 4. Results

### *4.1 SPH results*

In this section, we briefly summarize the results of the SPH impact simulations. A detailed description of the full set of simulations, as well as a discussion of other physical and numerical effects is given in *Emsenhuber et al. (in press)*.

*4.1.1 Material and temperature distribution resulting from the SPH simulations*

A snapshot of the grazing impact simulations at the time of the first contact of the impactor with the target is shown in figure 1 for the two different material rheologies. We note that with fluid rheology, the impactor is tidally deformed prior to the collision in a more pronounced way than with solid rheology. Qualitative differences in the collision outcome due to the equation of state and the rheology can be seen in figure 2, which shows for the two cases the material and temperature distributions (at $t = 18$ h after the impact) resulting from the impact. There are two main effects of using solid rheology observable in the final results: the impactor's material location and the heat distribution around the impact zone.

Considering material location, about 25 wt% of impactor's core remains in the upper part of the target's mantle under solid rheology whereas for fluid rheology this is insignificant. Also the distribution of the mantle of the impactor within the target is quite different. For solid rheology, about half of the impactor's mantle ends up in the interior while the remaining impactor mantle material is close to the surface. In contrast, with fluid rheology, about 10 wt% of the impactor's mantle lies close to the core-mantle boundary, very little across the rest of the target's mantle, and the majority is located close to the surface. Assuming solid rheology the heat distribution is focussed around the impact zone; in contrast, mostly the impactor material gets heated in models assuming a fluid rheology and the temperature distribution is less localised. This results in initially less silicate melt being present in all fluid rheology cases than in cases assuming a solid rheology.

The differences due to the different equations of state are more subtle. Material with high internal energy has a lower temperature with ANEOS compared to Tillotson. While ANEOS includes the latent heat of vaporisation, we compute the temperature from the specific energies in the runs with the Tillotson EoS using eq. (2) that does not take this into account. However, this affects only a small fraction of material, which is located close to the surface. This issue is discussed in more detail in *Emsenhuber et al., in press*. In future simulations with much higher surface resolution the effect may be significant.

*4.1.2 Re-impacting ejecta*

At the standard transfer time ($t = 18$ h) at which we switch from the SPH impact simulations to the interior evolution code I3ELVIS, there is still some ejecta material in orbit, which will potentially get accreted onto the target body later on. We use the following procedure to compute the amount of ejecta remaining to re-impact the Mars-sized target body as a function of time: For each SPH particle it is checked whether it is (1) bound to the target body, (2) on a crossing orbit and (3) outside a sphere with a radius of 3700 km around the center of mass of the target body. Figure 3 shows that within ~18 hours after the collision the remaining ejecta mass in orbit drops to values below $<10^{-3} M_{Mars}$.

To estimate the potential thermal effect of this remaining ejecta on the global evolution, we can calculate its thickness when distributed into a global surface layer of average thickness *d* by

$$d = \sqrt[3]{\frac{3(M+m)}{4\pi\bar{\rho}}} - R \quad (27)$$

where *M* is the mass of the target planet, *m* is the mass of the ejecta layer and $\bar{\rho}$ is the average density of the planet and *R* its average radius without the ejecta blanket. For simplicity we assume that the remaining orbiting ejecta and the planet have the same average density. However it should be kept in mind that local effects of falling ejecta not considered here may be important.

Using the characteristic diffusion time scale $t \sim d^2/\kappa$, where $\kappa$ is the thermal diffusivity, we can estimate the minimum cooling timescale of this remaining ejecta material (see Figure 4). For remaining ejecta masses $<10^{-3}M_{Mars}$ the characteristic diffusion timescale becomes very short compared to the crust formation timescale (see section 4.2), therefore we expect that the disregarded ejecta mass will not affect the evolution of the post-impact planet significantly.

*4.2 I3ELVIS results*

As described in section 3, for each impact simulation (see table 1) the SPH output is transferred into the I3ELVIS code used to study the post-impact interior evolution. In these longer-term calculations, the crust formation is computed by the procedure outlined in section 2.2.2. The resulting crustal distributions at $t \sim 5 \cdot 10^5$ yr after the collision are displayed in Figures 5-7. Since the early post-collision evolution involves the sinking of iron diapirs through the target's mantle, additional mantle melting occurs due to the release of potential energy. For the head-on collisions the bulk of the sinking iron is concentrated in one iron diapir, as a result most of the heat release occurs in the impact region. For the grazing collisions the post-impact iron distribution is more complex and various-sized iron diapirs are initially distributed throughout a larger portion of the target mantle (see Figure 2). Due to the power law rheology the sinking velocity of the iron diapirs is faster than in a Newtonian medium [e.g. *Weinberg & Podladchikov, 1994*]. Thus merging of the iron diapirs with the target's core occurs in all calculations on a timescale of ~$10^4$ yrs. Due to the shear heating related to the sinking of the iron diapirs hot wakes form [*Ziethe and Spohn, 2007*], evolving on the longer-term into thermal plumes. Due to the cooling of the silicate melt a basaltic crust forms, however delamination from the bottom occurs repeatedly. This delaminating crust sinks towards the core-mantle boundary and limits the crustal thickness. To determine the time-dependent total volume and distribution of crust, we consider only stable crustal material, meaning that disconnected crustal drips are disregarded. While we observe differences in the time evolution and the final volume of stable crust, all models show that the amount of stable crustal volume stabilizes after several hundred thousand years of evolution (see Figure 8). However, we observe a factor of two difference in the volume of stable crustal material for the different calculations, especially when considering the grazing collisions using Tillotson EoS with different rheologies.

Figure 5 shows the cumulative distribution of the crustal thickness for the head-on and grazing impacts at ~5· $10^5$ yr after the collision, respectively. A comparison of the results for the two different EoS is shown as well. In Figures 6 and 7, the corresponding crustal thickness maps are displayed, showing the spatial distribution for each case. Overall, the main differences in the spatial crustal pattern result from the impact angle. This effect can be clearly seen in the crustal thickness maps, which for head-on impacts show very different patters than for grazing impacts. In all cases considered, head-on impacts result in a hemispherical crustal dichotomy (see Figure 6). On the other hand the crustal pattern resulting from the grazing impacts is less consistent (see Figure 7).

As above, the differences between the runs with fluid and with solid rheology are generally more pronounced in the cases using Tillotson EoS (see Figure 5). These variations appear to mostly emerge from the different distributions of hot material just after the collision. As mentioned above a larger fraction of the silicates is molten in simulations with solid SPH rheology (coll02; see table 1), preferentially resulting in the formation of more crust in the long-term. However, the results for fluid rheology (coll08; see table 1) show at 300-1000 km depth the presence of hotter (~300 K) and more abundant (>5 times larger volume) impactor-derived melt in the mantle than in cases assuming a solid SPH rheology. For simulation coll02 this allows for fast cooling of the near surface layers and the crust tends to remain stable since no extended regions of still largely molten, thus low viscosity material, are present beneath the crust that would favour its destabilization. On the other hand, in simulation coll08 these regions of very hot silicate material are present for an extended time. Due to the presence of this low viscosity material beneath the newly formed crust the growth of crustal Rayleigh-Taylor instabilities is favoured. This is because the instability growth rate depends on the viscosity of the underlying material [*Turcotte & Schubert, 2014*]. This causes more crustal dripping events in simulation coll08 and more crustal material sinks into the mantle (see Figure 9). Comparing the two simulation outcomes it can be observed that only ~34% of the crustal material is stable at the end of simulation coll08 compared to ~56 % for simulation coll02.

For our nominal case (coll04; see table 1), we performed additional longer-term evolution simulations using different SPH-I3ELVIS transfer times. Figure 10 shows the corresponding crustal thickness distributions. These look relatively similar for the three cases (transfer time of 8 h, 12 h, 18 h) investigated, indicating that as expected the timing of the data transfer does not have a major effect on the crustal distribution.

Although this paper is only about the demonstration of the methodology, it is interesting to compare our results to the observed crustal volumes on a planet like Mars. At ~5· $10^5$ yr after the collision the volume of stable crust lies between 1.23· $10^{18}$ m$^3$ and 2.39· $10^{18}$ m$^3$, this corresponds to 17-34 % of the present-day volume of the martian crust [*Taylor & McLennan, 2009*]. Data indicate that at the end of the simulations 10-15% of the mantle material is partially molten and buoyant. This is related to the presence of hot plumes formed from the wakes left by the sinking iron diapirs. However most of the material in the shallow mantle is already highly depleted, therefore future crust formation is limited. Despite this some additional crust formation related to plumes can be expected to occur on the longer-term.

## 5. Discussion

The presented numerical models employing SPH and thermochemical simulations performed in series show that while the total volume of crust formed after a collision is comparable for both head-on and grazing collisions, the spatial crustal distribution is strongly affected by the impact angle. While we observe variations in the distribution and total crustal volume, the present results show that for both EoS and both SPH rheologies employed the resulting post-collision crust displays a crustal dichotomy, when considering a head-on collision, while this is not the case for all models assuming the more likely grazing collision.

Focussing on the outcome of the calculations based on SPH models with the more realistic solid rheology, the influence of the EoS on both the total volume and the spatial distribution of crustal material is more limited. The results also show that the rheology employed in the SPH models does have a significant influence on the immediate post-impact internal heat distribution and amount of silicate melt, but for the long-term crust formation process it is only a second-order effect. We note that its influence is more pronounced in the Tillotson cases. This might be related to the simplified treatment of silicate melting employed in models employing the Tillotson EoS. As discussed this can have an effect on the volume of stable crust since the presence of very hot material during the crust formation stage favours crustal dripping events.

### *5.1 Possible future improvements*

As mentioned above, some partial melt is still present at the end of the presented simulations. To study how much additional crust can be formed afterwards and whether all of the post-collision crust will be stable on the long-term additional calculations will be necessary in the future. A possible approach would be the transfer of the I3ELVIS results into the code StagYY [*Tackley, 2008*] suited to model the long-term evolution of terrestrial planets as done previously in 2-D geometry in *Golabek et al. [2011]*. Also it would be useful to study the effect of the impact angle on crust formation in more detail and determine where the transition between a dichotomous crust as observed for all head-on models and the more complex crustal pattern found for grazing collisions occurs in the parameter space. Also due to numerical limitations we have to use a lower cut-off for viscosity of all materials at $10^{18}$ Pa s. This limits the possibility of the post-collision magma ocean to spread due to isostatic relaxation [*Reese and Solomatov, 2006*]. A possible approach would require a reduced effective thermal conductivity, so magma ocean cooling would be extended, allowing for a longer spreading time. Additionally, a more realistic initial temperature profile with an increasing temperature with depth would allow for more efficient spreading since the viscosity of the deeper mantle would be reduced, thus isostatic relaxation could occur on a shorter timescale. Clearly these features have to be investigated in the future in more detail.

## 6. Conclusions

The present demonstration calculations suggest that the coupled SPH-thermochemical approach in 3-D geometry, although computationally expensive, can help in the future to obtain a more realistic post-collision evolution of planetary interiors as these models are able to capture both the collision process, the following magma ocean cooling, crust formation and solid-state deformation processes in detail. Our results indicate that significant amounts of crustal material taking a dichotomous crustal pattern can form after a head-on collision, while this is not the case when considering a more likely grazing collision. This coupled approach can be applied to various planetary bodies to improve our understanding of their early evolution.

## Acknowledgements

We thank two anonymous reviewers for detailed comments that helped to improve the manuscript considerably. Helpful discussions with Francis Nimmo and Don Korycansky on giant impacts are appreciated. A.E. acknowledges the financial support of the Swiss National Science Foundation under grant 200020_17246. This work has been carried out within the frame of the National Centre for Competence in Research PlanetS supported by the Swiss National Science Foundation.

**References**

Alemi, A., Stevenson, D.J., 2006. Why Venus has No Moon, American Astronomical Society, DPS meeting #38, #07.03; *Bulletin of the American Astronomical Society*, 38, 491.

Andrews-Hanna, J.C., Zuber, M.T., Banerdt, W.B., 2008. The Borealis basin and the origin of the martian crustal dichotomy. *Nature* 453, 1212-1215.

Asphaug, E., 2010. Similar-sized collisions and the diversity of planets. *Chem. Erde - Geochemistry* 70, 199-219.

Asphaug, E. & Reufer, A., 2014. Mercury and other iron-rich planetary bodies as relics of inefficient accretion. *Nature Geosci.* 7, 564-568.

Benz, W., 1991. An Introduction to Computation Methods in Hydrodynamics. In: Loore, C.B. de, (Ed.), *Lecture Notes in Physics* 373, Late stages of stellar evolution, computational methods in astrophysical hydrodynamics. Springer, pp. 258-312.

Benz, W., Cameron, A.G.W., Slattery, W.L., 1988. Collisional Stripping of Mercury's Mantle. *Icarus* 74, 516-528.

Benz, W., Asphaug, E., 1994. Impact simulations with fracture. I - Method and tests. *Icarus* 107, 98-116.

Benz, W., Anic, A., Horner, J., Whitby, J.A., 2007. The Origin of Mercury. *Space Sci. Rev.* 132, 189-202.

Bottinga, Y., Weill, D.F., 1972. The viscosity of magmatic silicate liquids: A model for calculation. *Am. J. Sci.* 272, 438-475.

Burg, J.-P., Gerya, T.V., 2005. The role of viscous heating in Barrovian metamorphism of collisional orogens: Thermomechanical models and application to the Lepontine Dome in the Central Alps. J. Metamorp. Geol. 23, 75-95.

Cameron, A.G.W., Ward, W.R., 1976. The origin of the Moon. *Lunar Sci.* 7, 120-122.

Canup, R.M., Asphaug, E., 2001. Origin of the Moon in a giant impact near the end of the Earth's formation. *Nature*, 412, 708-712.

Canup, R.M., 2004. Simulations of a late lunar-forming impact. *Icarus* 168, 433-456.

Canup, R.M., 2012. Forming a Moon with an Earth-like Composition via a Giant Impact. *Science* 338, 1052-1055.

Canup, R.M., Barr, A.C., Crawford, D.A. 2013. Lunar-forming impacts: High-resolution SPH and AMR-CTH simulations. *Icarus* 222, 200-219.

Collins, G.S., Melosh, H.J., Ivanov, B.A., 2004. Modeling damage and deformation in impact simulations. *Meteor. Planet. Sci.* 39, 217-231.

Condomines, M., Hemond, C., Allègre, C.J., 1988. U-Th-Ra Radioactive Disequilibria and Magmatic Processes. *Earth Planet. Sci. Lett.* 90, 243-262.

Ćuk, M., Stewart, S.T., 2012. Making the Moon from a Fast-Spinning Earth: A Giant Impact Followed by Resonant Despinning. *Science* 338, 1047-1052, 2012.

Chudinovskikh, L., Boehler, R., 2007. Eutectic melting in the system Fe-S to 44 GPa. *Earth Planet. Sci. Lett.* 257, 97-103.

Costa, A., Caricchi, L., Bagdassarov, N., 2009. A model for the rheology of particle-bearing suspensions and partially molten rocks. *Geochem. Geophys. Geosyst.* 10, Q03010.

Crameri, F., Schmeling, H., Golabek, G.J., Duretz, T., Orendt, R., Buiter, S.J.H., May, D.A., Kaus, B.J.P., Gerya T.V., Tackley, P.J., 2012. A comparison of numerical surface topography calculations in geodynamic modelling: an evaluation of the 'sticky air' method. *Geophys. J. Int.* 189, 38-54.

Emsenhuber, E., Jutzi, M., Benz, W., SPH calculations of Mars-scale collisions: The role of the equation of state, material rheologies, and numerical effects. *Icarus*, in press.

Escartín, J., Hirth, G., Evans, B., 2001. Strength of slightly serpentinized peridotites: Implications for the tectonics of oceanic lithosphere. *Geology* 29, 1023-1026.

Frey, H., Schultz, R.A., 1988. Large impact basins and the mega-impact origin for the crustal dichotomy on Mars. *Geophys. Res. Lett.* 15, 229-232.

Gerya, T.V., Yuen, D.A., 2003. Characteristics-based marker-in-cell method with conservative finite-differences schemes for modeling geological flows with strongly variable transport properties. *Phys. Earth Planet. Int.* 140, 293-318.

Gerya, T.V., Yuen, D.A., 2007. Robust characteristics method for modelling multiphase visco-elasto-plastic thermo-mechanical problems. *Phys. Earth Planet. Int.* 163, 83-105.

Gilbert, G.K., 1893. The Moon's face, a study of the origin of its features. *Bull. Philos. Soc. Wash. (DC)* 12, 241-292.

Golabek, G.J., Keller, T., Gerya, T.V., Zhu, G., Tackley, P. J., Connolly, J.A.D., 2011. Origin of the martian dichotomy and Tharsis from a giant impact causing massive magmatism. *Icarus* 215, 346-357.

Golabek, G.J., Bourdon, B., Gerya, T.V., 2014. Numerical models of the thermomechanical evolution of planetesimals: Application to the acapulcoite-lodranite parent body. *Meteorit. Planet. Sci.* 49, 1083-1099.

Hartmann, W. K., Davis, D. R., 1975. Satellite-sized planetesimals and lunar origin. *Icarus* 24, 504-515.

Hevey, P.J., Sanders, I.S., 2006. A model for planetesimal meltdown by $^{26}$Al and its implications for meteorite parent bodies, *Meteorit. Planet. Sci.* 41, 95-106.

Herzberg, C., Raterron, P., Zhang, J., 2000. New experimental observations on the anhydrous solidus for peridotite KLB-1. *Geochem. Geophys. Geosyst.* 1, 2000GC000089.

Jacobsen, B., Yin, Q.-Z., Moynier, F., Amelin, Y., Krot, A.N., Nagashima, K., Hutcheon, I.D., Palme, H., 2008. $^{26}$Al-$^{26}$Mg and $^{207}$Pb-$^{206}$Pb systematics of Allende CAIs: Canonical solar initial $^{26}$Al/$^{27}$Al ratio reinstated. *Earth Planet. Sci. Lett.* 272, 353-364.

Jutzi, M., Asphaug, E., Gillet, P., Barrat, J.-A., Benz, W., 2013. The structure of the asteroid 4 Vesta as revealed by models of planet-scale collisions. *Nature* 494, 207-210.

Jutzi, M., 2015. SPH calculations of asteroid disruptions: The role of pressure dependent failure models. *Planet. Space Sci.* 107, 3-9.

Jutzi, M., Holsapple, K.A., Wünnemann, K., Michel, P., 2015. Modeling asteroid collisions and impact processes. In *Asteroids IV* (P. Michel, F. DeMeo, and W. F. Bottke, eds.), Univ. of Arizona, Tucson, pp. 679-700.

Kameyama, M., Yuen, D.A., Karato, S.-i., 1999. Thermal-mechanical effects of low-temperature plasticity (the Peierls mechanism) on the deformation of a viscoelastic shear zone, *Earth Planet. Sci. Lett.* 168, 159-172.

Karato, S.-i., Murthy, V.R., 1997. Core formation and chemical equilibrium in the Earth I. Physical considerations. *Phys. Earth Planet. Int.* 100, 61-79.

Katayama, I., Karato, S.-i., 2008. Low-temperature, high-stress deformation of olivine under water-saturated conditions. *Phys. Earth Planet. Int.* 168, 125-133.

Keller, T., Tackley, P.J., 2009. Towards self-consistent modelling of the martian dichotomy: The influence of one-ridge convection on crustal thickness distribution. *Icarus* 202, 429-443.

Kraichnan R.H., 1962. Turbulent thermal convection at arbitrary Prandtl number. *Phys. Fluids* 5, 1374-1389.

Liebske, C., Schmickler, B., Terasaki, H., Poe, B.T., Suzuki, A., Funakoshi, K.-I., Ando, R., Rubie, D.C., 2005. Viscosity of peridotite liquid up to 13 GPa: Implications for magma ocean viscosities. *Earth Planet. Sci. Lett.* 240, 589-604.

Lodders, K., Fegley, B., 1998. The Planetary Scientist's Companion. Oxford Univ. Press, New York, 371pp.

Mackwell, S.J., 1991. High-temperature rheology of enstatite: Implications for creep in the mantle. *Geophys. Res. Lett.* 18, 2027-2030.

Marinova, M.M., Aharonson, O., Asphaug, E., 2008. Mega-impact formation of the Mars hemispheric dichotomy. *Nature* 453, 1216-1219.

Marinova, M.M., Aharonson, O., Asphaug, E., 2011. Geophysical consequences of planetary scale impacts into a Mars-like planet. *Icarus* 211, 960-985.

Melosh, H.J., 1989. Impact Cratering: A Geologic Process. Monographs on Geology and Geophysics. 11.

Melosh, H.J., 1990. Giant impacts and the thermal state of the early Earth. In: *Origin of the Earth*, edited by H. E. Newsom & J. H. Jones, Oxford Univ. Press, 69-83.

Melosh, H.J., 2007. A hydrocode equation of state for $SiO_2$. *Meteor. Planet. Sci.* 42, 2079-2098.

McKenzie, D., Bickle, M.J., 1988. The volume and composition of melt generated by extension of the lithosphere. *J. Petr.* 29, 625-679.

Mezger, K., Debaille, V., Kleine, T., 2013. Core Formation and Mantle Differentiation on Mars. *Space Science Rev.* 174, 27-48.

Morishima, R., Golabek, G.J. and Samuel, H., 2013. N-body simulations of oligarchic growth of Mars: Implications for Hf-W chronology. *Earth Planet. Sci. Lett.* 366, 6-16.

Nimmo, F., Kleine, T., 2007. How rapidly did Mars accrete? Uncertainties in the Hf-W timing of core formation. *Icarus* 191, 497-504.

Nimmo, F., Hart, S.D., Korycansky, D.G., Agnor, C.B., 2008. Implications of an impact origin for the martian hemispheric dichotomy. *Nature* 453, 1220-1223.

Pinkerton, H., Stevenson, R.J., 1992. Methods of determining the rheological properties of magmas at subliquidus temperatures. *J. Volcan. Geotherm. Res.* 53, 47-66.

Ranalli, G., 1995. Rheology of the Earth, second ed., Chapman & Hall, London, UK, 436pp.

Reese, C.C., Solomatov, V.S., 2006. Fluid Dynamics of local martian magma oceans. *Icarus* 184, 102-120.

Reese, C.C., Solomatov, V.S., 2010. Early martian dynamo generation due to giant impacts. *Icarus* 207, 82-97.

Reese, C.C., Orth, C.P., Solomatov, V.S., 2010. Impact origin for the martian crustal dichotomy: Half-emptied or half-filled? *J. Geophys. Res.* 115, E05004.

Reufer, A., Meier, M.M.M., Benz, W. & Wieler, R., 2012. A hit-and-run Giant Impact scenario. *Icarus* 221, 296-299.

Roberts, J.H., Zhong, S., 2006. Degree-1 convection in the Martian mantle and the origin of the hemispheric dichotomy. J. Geophys. Res. 111, E06013.

Rolf, T., Zhu, M.-H., Wünnemann, K., Werner, S.C., 2017. The role of impact bombardment history in lunar evolution. *Icarus* 286, 138-152.

Rubie, D., Melosh, H.J., Reid, J.E., Liebske, C., Righter, K., 2003. Mechanisms of metal-silicate equilibration

in the terrestrial magma ocean. *Earth Planet. Sci. Lett.* 205, 239-255.

Rubie, D.C., Nimmo, F., Melosh, H.J., 2015. Formation of the Earth's core. In Treatise on Geophysics, 2nd ed., Volume 9: Evolution of the Earth. Stevenson D. J. (ed.), Amsterdam: Elsevier B.V. pp. 43-79.

Schmeling, H., Babeyko, A.Y., Enns, A., Faccenna, C., Funiciello, F., Gerya, T.V., Golabek, G.J., Grigull, S., Kaus, B.J.P., Morra, G., Schmalholz, S.M., van Hunen, J., 2008. A benchmark comparison of spontaneous subduction models - Towards a free surface. *Phys. Earth Planet. Int.* 171, 198-223.

Shoemaker, E.M., 1962. Interpretation of lunar craters. In: Kopal, Z. (Ed.), *Physics and Astronomy of the Moon.* Academic Press, New York, pp. 283-359.

Siggia, E.D., 1994. High Rayleigh number convection. Annu. Rev. Fluid Mech. 26, 137-168.

Solomatov, V.S., 2015. Magma oceans and primordial mantle differentiation. In: Stevenson, D.J. (Ed.), Treatise on Geophysics, 2nd ed., Evolution of the Earth, vol. 9., Elsevier B.V., Amsterdam, Netherlands, pp. 81-104.

Senft, L.E., Stewart, S.T., 2009. Dynamic fault weakening and the formation of large impact craters. *Earth Planet. Sci. Lett.* 287, 471-482.

Šrámek, O., Zhong, S., 2010. Long-wavelength stagnant-lid convection with hemispheric variation in lithospheric thickness: link between Martian crustal dichotomy and Tharsis? *J. Geophys. Res.* 115, E09010.

Šrámek, O., Zhong, S., 2012. Martian crustal dichotomy and Tharsis formation by partial melting coupled to early plume migration. *J. Geophys. Res.* 117, E01005.

Stolper, E., Hager, B.H., Walker, D., Hays, J.F., 1981. Melt segregation from partially molten source regions - The importance of melt density and source region size. *J. Geophys. Res.* 86, 6261-6271.

Suzuki, A., Ohtani, E., Kato, T., 1998. Density and thermal expansion of a peridotite melt at high pressure. *Phys. Earth Planet. Int.* 107, 53-61.

Tackley, P.J., Schubert, G., Glatzmaier, G.A., Schenk, P., Ratcliff, J.T., 2001. Three-dimensional simulations of mantle convection in Io. *Icarus* 149, 79-93.

Tackley, P.J., 2008. Modelling compressible mantle convection with large viscosity contrasts in a three-dimensional spherical shell using the yin-yang grid. *Phys. Earth Planet. Int.* 171, 7-18.

Taylor, S.R., McLennan, S.M., 2009. Planetary Crusts: Their Composition, Origin and Evolution. Cambridge Univ. Press, Cambridge, UK, 404pp.

Thompson, S.L., Lauson, H.S., 1972. Improvements in the CHART-D Radiation-hydrodynamic code III: Revised analytic equations of state. *Technical report SC-RR-71 0714*.

Thrane, K., Bizzarro, M., Baker, J.A., 2006. Extremely brief formation interval for refractory inclusions and uniform distribution of $^{26}$Al in the early Solar System. *Astrophys. J.* 646, L159–L162.

Tillotson, J.H., 1962. Metallic Equations of State for Hypervelocity Impact. *GA-3216, General Atomic, San Diego, CA*.

Tkalcec, B.J., Golabek, G.J., Brenker, F.E., 2013. Solid-state plastic deformation in the dynamic interior of a differentiated asteroid. *Nat. Geosci.* 6, 93-97.

Trønnes, R.G., Frost, D.J., 2002. Peridotite melting and mineral-melt partitioning of major and minor elements at 22–24.5 GPa. *Earth Planet. Sci. Lett.* 197, 117-131.

Turcotte D. L., Schubert G., 2014. Geodynamics, 3rd ed. New York: Cambridge University Press. 636 pp.

Wade, J., Wood, B.J., 2005. Core formation and the oxidation state of the Earth. *Earth Planet. Sci. Lett.* 236, 78-95.

Weinberg, R.F., Podladchikov, Y., 1994. Diapiric ascent of magmas through power law crust and mantle. *J. Geophys. Res.* 99, 9543-9559.

Weinstein, S.A., 1995. The effects of a deep mantle endothermic phase change on the structure of thermal convection in silicate planets. *J. Geophys. Res.* 100, 11719-11728.

Wetherill, G.W., 1985. Occurrence of giant impacts during the growth of the terrestrial planets. *Science* 228, 877-879.

Wilhelms, D.E., Squyres, S.W., 1984. The martian dichotomy may be due to a giant impact. *Nature* 309, 138-140.

Yunker, M.L., Van Orman, J.A., 2007. Interdiffusion of solid iron and nickel at high pressure, *Earth Planet. Sci. Lett.* 254, 203-213.

Zahnle, K.J., Kasting, J.F., Pollack, J.B., 1988. Evolution of a steam atmosphere during Earth's accretion. *Icarus* 74, 62-97.

Ziethe, R., Spohn, T., 2007. Two-dimensional stokes flow around a heated cylinder: A possible application for diapirs in the mantle. *J. Geophys. Res.* 112, B09403.

Zhong, S., 2009. Migration of Tharsis volcanism on Mars caused by differential rotation of the lithosphere. *Nat. Geosci.* 2, 19-23.

Zhong, S., Zuber, M.T., 2001. Degree-1 mantle convection and the crustal dichotomy on Mars. *Earth Planet. Sci. Lett.* 189, 75-84.

**Figures**

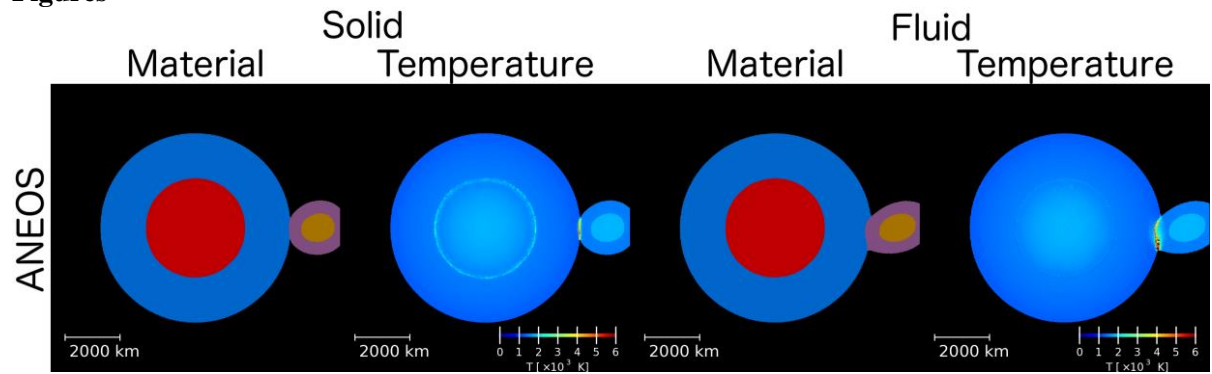

*Figure 1: Grazing impact simulations at t = 0 s. The impactor's trajectory is clockwise. Shown is a slice at 1000 km depth inside the impact plane. Plots are centered on the center of mass of the target body. Only simulations using ANEOS are shown here; models using Tillotson show almost no difference at this stage. Temperature distributions for solid and fluid rheologies according to captions. Colors for material distribution are as follows: blue for target mantle, purple for impactor mantle, red for target core and orange for impactor core.*

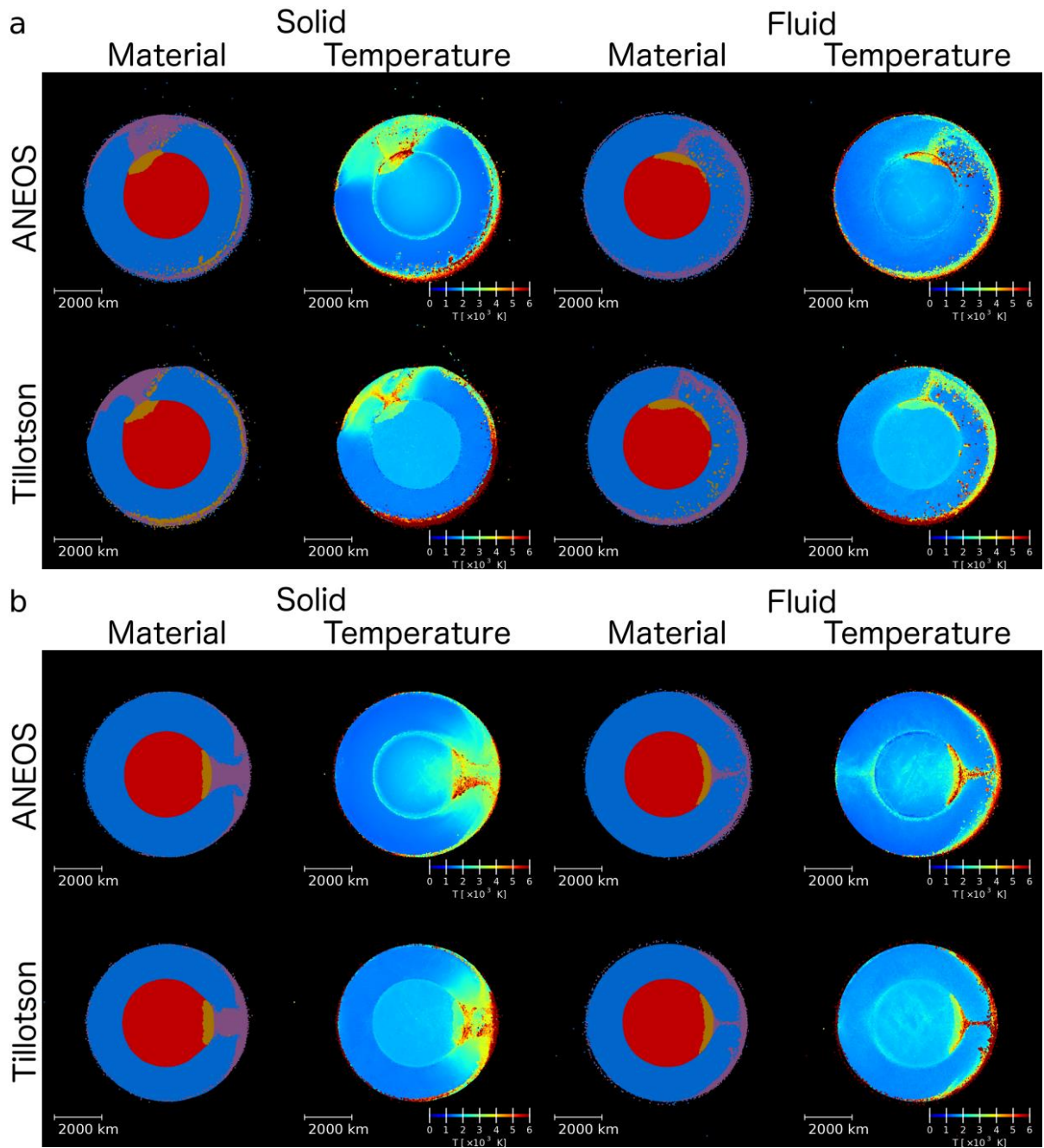

*Figure 2: Material and temperature distributions resulting from the (a) grazing collision simulations and (b) head-on collisions, at t = 18 hours after the impact. Colors for material distribution as given in Figure 1. Note that due to the grazing impacts, the targets have rotated clockwise by ~270 degrees.*

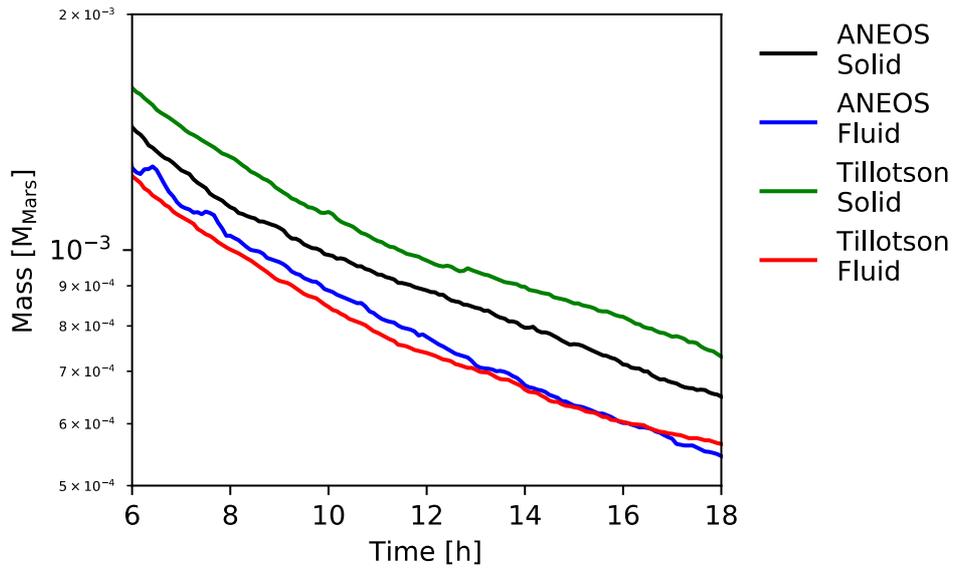

*Figure 3: Remaining ejecta in orbit as a function of time for the grazing cases shown in Figure 2a.*

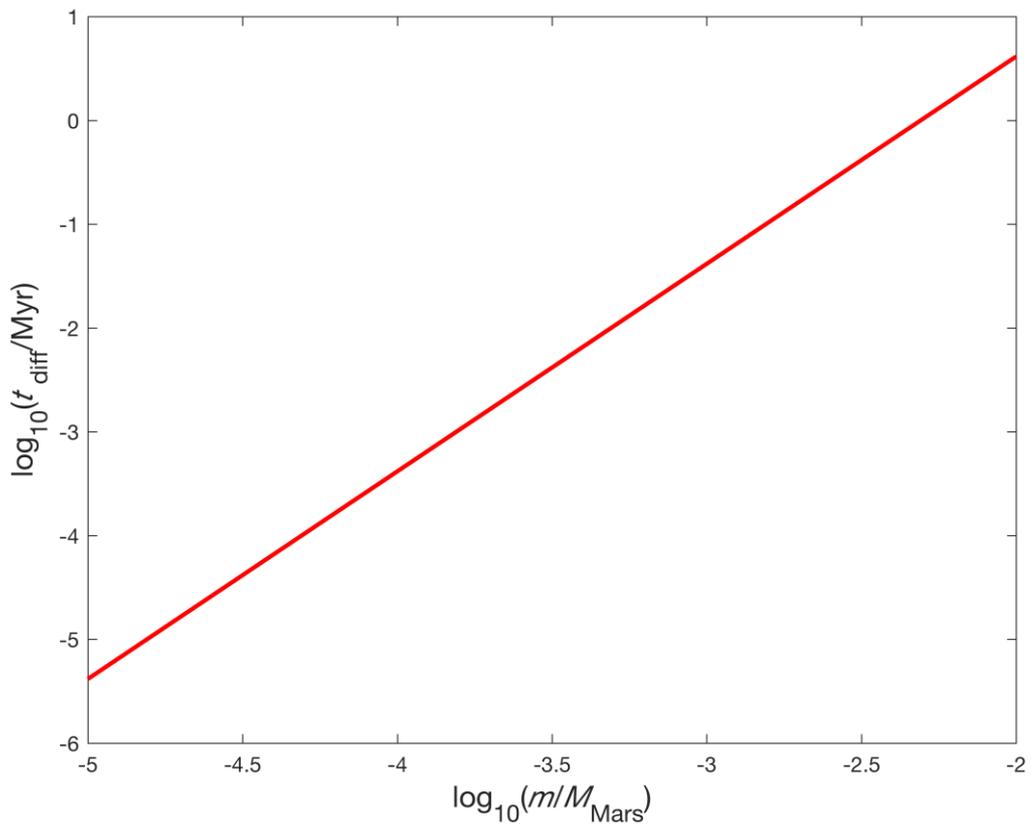

*Figure 4: Characteristic cooling timescale for the disregarded ejecta material when spread out into a global surface layer.*

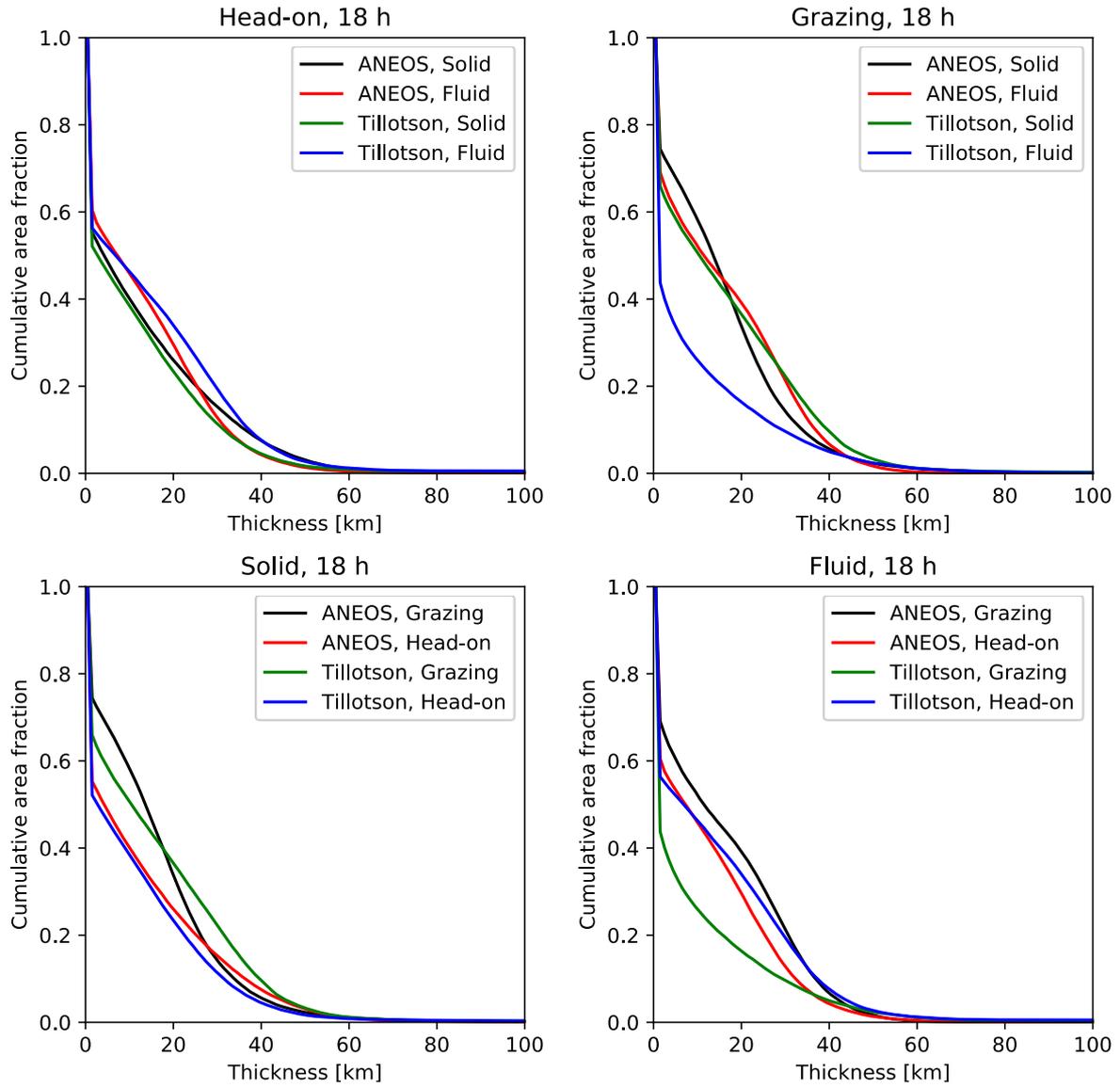

*Figure 5: Cumulative crustal thickness distributions at ~5·10$^5$ yr after the collision for different impact angle (upper row) and different rheology used in the SPH simulation (lower row). The curves show for each case the fraction of the surface area that has an underlying crust of a thickness larger than the value given on the x-axis.*

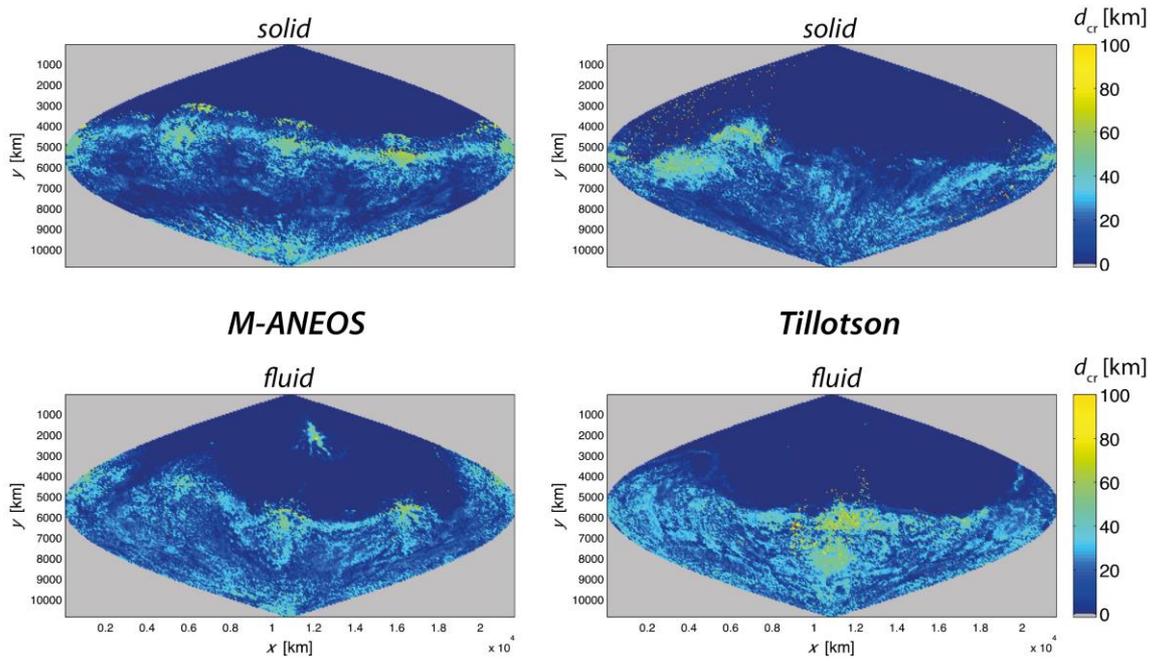

*Figure 6: Maps of crustal thickness distribution at ~5·10$^5$ yr after the collision resulting from head-on collisions using the Mercator equal-area projection.*

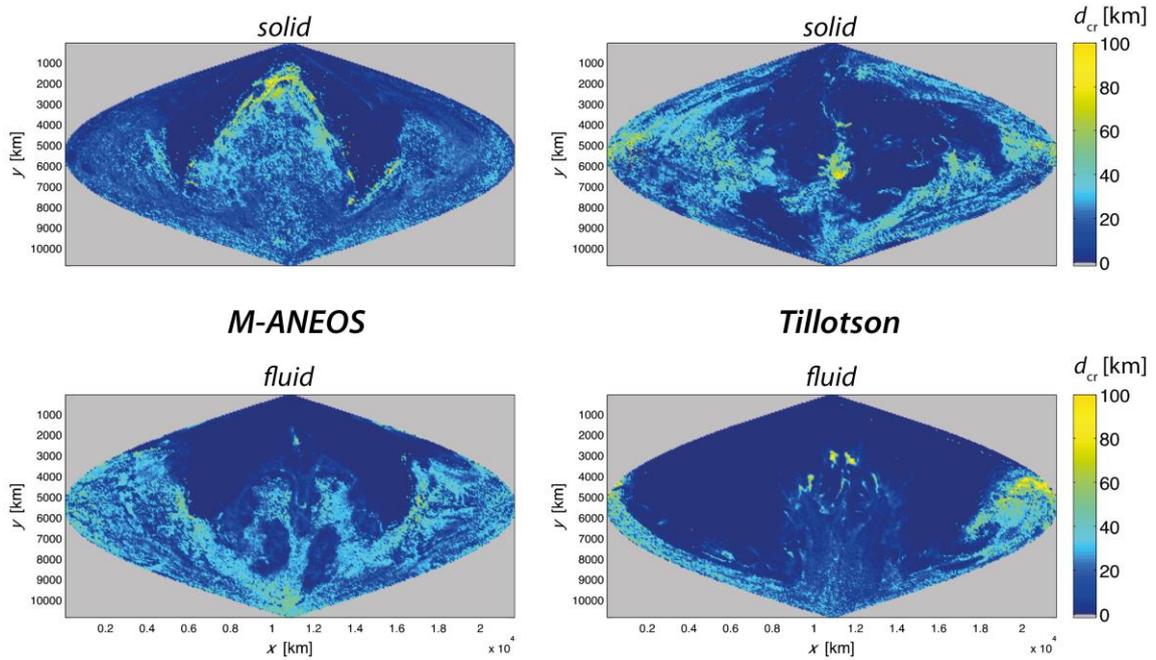

*Figure 7: Maps of crustal thickness distribution at ~5·10$^5$ yr after the collision resulting from grazing collisions using the Mercator equal-area projection.*

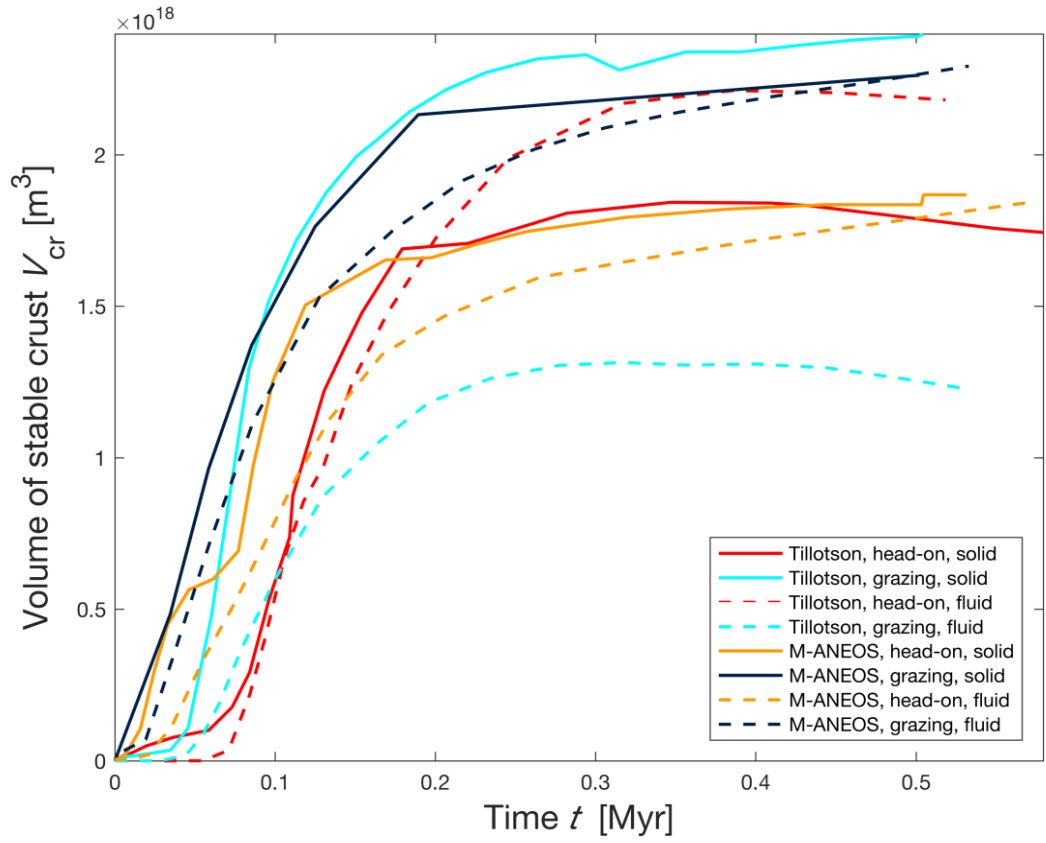

*Figure 8: Time evolution of the volume of stable crust (disconnected crustal drips are disregarded).*

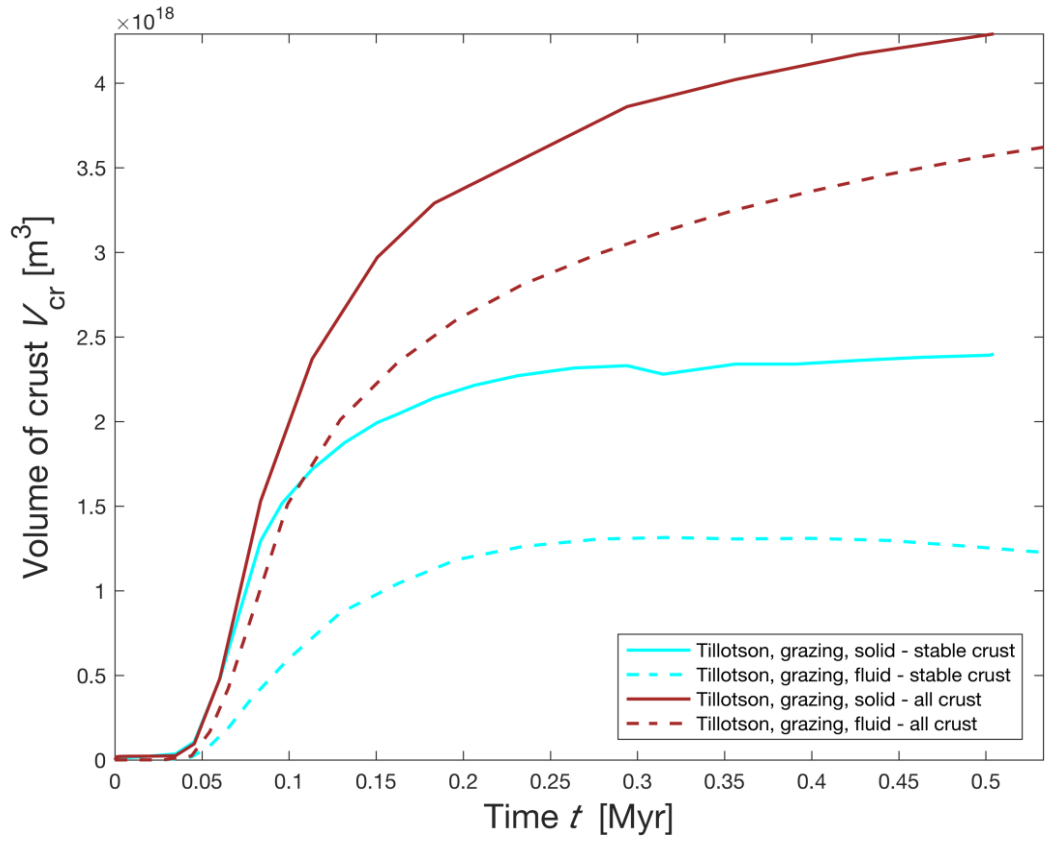

*Figure 9: Time evolution of the volume of stable crust (cyan coloured lines) and total volume of crust (auburn coloured lines) for the two extreme cases (coll02, coll08) from Figure 8.*

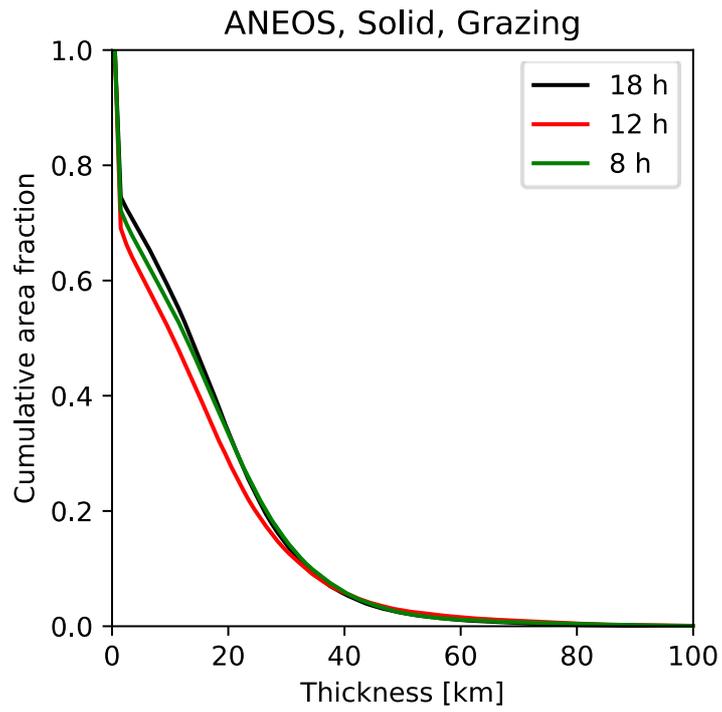

*Figure 10: Cumulative crustal thickness distributions at ~5·10$^5$ yr after the collision for the nominal case using different SPH-I3ELVIS transfer times.*

**Tables**

*Table 1: List of numerical models*

| Model name | Impact angle [degrees] | EoS | Transfer time SPH-I3ELVIS [h] | SPH rheology |
|---|---|---|---|---|
| coll01 | 0 | Tillotson | 18 | solid |
| coll02 | 45 | Tillotson | 18 | solid |
| coll03 | 0 | M-ANEOS | 18 | solid |
| coll04 | 45 | M-ANEOS | 18 | solid |
| coll05 | 45 | M-ANEOS | 8 | solid |
| coll06 | 45 | M-ANEOS | 12 | solid |
| coll07 | 0 | Tillotson | 18 | fluid |
| coll08 | 45 | Tillotson | 18 | fluid |
| coll09 | 0 | M-ANEOS | 18 | fluid |
| coll10 | 45 | M-ANEOS | 18 | fluid |

*Table 2: Physical parameters used in I3ELVIS*

| Parameter | Symbol | Value | Units | Reference |
|---|---|---|---|---|
| Density of uncompressed silicate melt | $\rho_{\text{Si-liq}}$ | 2900 | kg/m$^3$ | (1,2) |
| Density of uncompressed solid silicates | $\rho_{\text{Si-sol}}$ | 3500 | kg/m$^3$ | (1) |
| Temperature of space (sticky air) | $T_{\text{sa}}$ | 220 | K | (3) |
| Activation energy | $E_{\text{a}}$ | 470 | kJ/mol | (4) |
| Activation volume | $V_{\text{a}}$ | 8·10$^{-6}$ | m$^3$/mol | (5) |
| Dislocation creep onset stress | $\sigma_0$ | 3·10$^7$ | Pa | (6) |
| Power law exponent | $n$ | 4 |  | (4) |
| Cohesion | $C$ | 10$^8$ | Pa | (5) |
| Internal friction coefficient of solid silicates | $f$ | 0.3 |  | (7) |
| Latent heat of silicate melting | $L_{\text{Si}}$ | 400 | kJ/kg | (6) |
| Silicate melt fraction at rheological transition | $\varphi_{\text{crit}}$ | 0.4 |  | (8,9) |
| Heat capacity of silicates | $c_{\text{P}}$ | 1000 | J/(kg K) | (6) |
| Thermal expansivity of solid silicates | $\alpha_{\text{Si-sol}}$ | 3·10$^{-5}$ | 1/K | (2) |
| Thermal expansivity of molten silicates | $\alpha_{\text{Si-liq}}$ | 6·10$^{-5}$ | 1/K | (2) |
| Thermal conductivity of solid silicates | $k$ | 3 | W/(m K) | (6) |
| Thermal conductivity of molten silicates | $k_{\text{eff}}$ | ≤10$^6$ | W/(m K) | (10) |

References: (1) Stolper et al. (1981), (2) Suzuki et al. (1998), (3) Lodders & Fegley (1998), (4) Ranalli (1995), (5) Golabek et al. (2011), (6) Turcotte & Schubert (2014), (7) Escartín et al. (2001), (8) Solomatov (2015), (9) Costa et al. (2009), (10) Golabek et al. (2014)